\documentclass[showpacs,preprintnumbers,amsmath,amssymb,aps]{revtex4}

\usepackage{graphics}

\usepackage{dcolumn}
\usepackage{bm}

\begin{document}
\title{Modified Newtonian Dynamics and Induced gravity}
\author{
W. F. Kao\thanks{%
gore@mail.nctu.edu.tw} \\
Institute of Physics, Chiao Tung University, Hsinchu, Taiwan}

\begin{abstract}
Modified Newtonian dynamics, a successful alternative to the
cosmic dark matter model, proposes that gravitational field
deviates from the Newtonian law when the field strength $g$ is
weaker than a critical value $g_0$. We will show that the dynamics
of MOND can be derived from an induced gravity model. New dynamics
is shown to be compatible with the spatial deformation of scalar
fields coupled to the system. Approximate solutions are shown
explicitly for a simple toy model.
\end{abstract}


\maketitle

\section{INTRODUCTION}
 The rotation curve (RC) observations indicates that less than 10
\% of the gravitational mass can be measured from the luminous
part of spiral galaxies. This is the first evidence calling for
the existence of un-known dark matter and dark energy. In the
meantime, an alternative approach, Modified Newtonian dynamics
(MOND) proposed by Milgrom \cite{milgrom}, has been shown to agree
with many rotation curve observations \cite{milgrom, sanders}.

Milgrom argues that dark matter is redundant in the approach of
MOND. The missing part was, instead, proposed to be derived from
the conjecture that gravitational field deviates from the
Newtonian $1/r^2$ form when the field strength $g$ is weaker than
a critical value $g_0 \sim 0.9 \times 10^{-8} $ cm/s$^2$
\cite{sanders}.

The phenomenological foundations for MOND are based on two
observations: (1) flat asymptotic rotation curve (RC) is a common
feature for many spiral galaxies, (2) the Tully-Fisher (TF) law,
$M \sim V^\alpha$ \cite{tf77} is very successful explaining the
relation between rotation velocity and luminosity in many spiral
galaxies. $\alpha$ is shown to be close to 4. Note that first fact
indicates that gravitational field $g$ goes like $1/r$
asymptotically in the flat RC region.

There have been effort trying to connect non-relativistic MOND
theory with a relativistic version\cite{bek}. We will show that
the dynamics of MOND could be due to the effect of scalar fields
in an induced gravity model. It is known that scalar field can be
used to induce mass for a massless fermion from a Yukawa coupling.
The associated effective potential is also a source of dark
energy. In fact, one of the original effect of the scalar field is
to deform the definition of scale and distance in the Weyl
invariant model.

Indeed, Weyl \cite{weyl, cheng} proposed that the invariant length
scale should be defined as $ds^2 \equiv \phi g_{\mu \nu} dx^\mu
dx^\nu$. It is easy to find that $ds^2$ is invariant under the
local scale transformation $\phi(x) \to \Lambda^2(x) \phi(x)$ and
$g_{\mu \nu} \to \Lambda^{-2}(x)g_{\mu\nu}$. The local length
scale $d\tau^2 \equiv g_{\mu \nu} dx^\mu dx^\nu$ measured by any
observers will then be deformed due to the scale chosen by the
dimension two scalar field $\phi$.

Therefore one of the effect of the scalar field is to deform the
effective distance between two distant matters. One will show
explicitly in this paper a possible resolution derived from a
simple model with a spherically symmetric mass-distribution
directed by similar effect.

Einstein theory works with the latest CMB anisotropic survey in a
manageable way. One would like to study if there is any way to
induce the MOND under the Einstein-Hilbert framework with minimum
modifications. Indeed, one is able to show that the dynamics of
MOND could be due to the large-scale spatial inhomogeneity of the
coupled scalar field. A specific example will be presented in this
paper to show its effect in the large $r$ limit where the
effective potential $V(r)$ goes like $\ln r$ in the absence of any
dark matter.

In section II, the idea of MOND will be briefly reviewed. The
force law of MOND will also be rewritten in term of dimensionless
field variables for convenience in this section. The effective
potential for MOND will be shown in section III. In section IV,
one shows that induced gravity model could provide a possible
resolution to the force law of MOND with the help of two scalar
fields $\psi$ and $\phi$. Two different approximate solutions to
the field equations associated with a simple toy model are
presented in section V. Some conclusions and remarks are drawn in
section VI.

\section{MOND}

It was proposed that there exists a critical acceleration
parameter $g_0 = 0.9 \times 10^{-8} {\rm cm s}^{-2} $
\cite{sanders} (Sanders 2002) characterizing the turning point of
the effective power law associated with the gravitational field in
MOND. Gravitational field of the following form was suggested
\begin{equation} \label{gm}
g \cdot \mu ({g \over g_0}) =g_N
\end{equation}
with a function $\mu(x)$ considered as modified inertial. Here
$g_N$ is the Newtonian gravitational field produced by any sort of
mass distribution. Milgrom suggests that the following
function-form
\begin{equation}
\mu (x) = {x  \over  \sqrt{1+x^2}}
\end{equation}
agrees with the rotational curve of many spiral galaxies.

It is interesting to find that there exists a length scale factor
$r_0$ associated with our Milky Way via the following equation
\begin{equation} \label{g0}
g_0 \equiv {G M_0 \over r_0^2}
\end{equation}
with $M_0 \equiv 2 \times 10^{11} M_s$ roughly the order of total
mass of the Milky Way. $M_s$ denotes the solar mass. As a result
one can show that $r_0 \sim 49700 ly$ which is roughly the size of
the luminous part our Milky Way.

Note that $M_0$ is the typical mass of any galaxy with a size
similar to our Milky Way. In fact, $r_0$ chosen here will only
served as a convenient unit of length scale . Selecting different
value of $M_0$ and hence $r_0$ will not affect the physics
presented in this paper. It turns out that writing all physical
variables in dimensionless form will make it easier for us to
focus on important physics.

One will try to write the modified field strength $g$ in a
dimensionless form with the help of the parameters $M_0$, $r_0$
and $g_0$. The idea is to write $M$, $r$ and $g$ in units of
$M_0$, $r_0$ and $g_0$ respectively.

Milgrom argues that one can take the effective inertial $\mu$ as
$\mu (x) = x / \sqrt{1+x^2} $. Therefore, one obtains
\begin{equation} \label{g}
{(g/g_0)^2 \over \sqrt{1+(g/g_0)^2} }= g_N/g_0
\end{equation}
by dividing both sides of the equation [\ref{gm}] by the critical
parameter $g_0$. One can therefore write above equation as
\begin{equation}
{ g^2 \over \sqrt{1+g^2} }= g_N
\end{equation}
with $g$ and $g_N$ written in unit of $g_0$. This makes $g$ and
$g_N$ dimensionless from now on.

Throughout this paper, we will focuss on the study of the system
with a Newtonian attraction of the form $g_N =Gm/r^2$ for
simplicity. This is the field strength at a radial distance $r$
from a spherically distributed matter with total mass $m$. System
with different mass distribution can be obtained
straightforwardly. To summarize, one has chosen dimensionless
scale according to following redefinitions:
\begin{eqnarray} \label{dim}
m' &=& {m \over M_0}, \nonumber \\
g' &=& {g \over g_0}  , \nonumber \\
r' &=& { r \over r_0}.
\end{eqnarray}
For example, $g'_N=m'/r'^2$. One will suppress the superscript $'$
for convenience. Therefore, the modified field strength $g$
becomes
\begin{equation}
{ g^2 \over \sqrt{1+g^2} }=  {m \over r^2} \label{01}
\end{equation}
One can further remove the parameter $m$ by defining $r_c \equiv
r_0 \sqrt{m}$ and write $r$ as a dimensionless coordinate variable
in unit of $r_c$ instead of $r_0$.
\begin{eqnarray} \label{dim0}
r'' ={r \over r_c}={ r \over \sqrt{m}r_0}
\end{eqnarray}
As a result, one can write
above equation in a very compact form
\begin{equation}
{ g^2 \over \sqrt{1+g^2} }=  {1 \over r''^2} \label{9}
\end{equation}
in terms of the dimensionless parameters $g$, $g_N$, and $r''$.
Note that it is straightforward to restore all dimension
parameters to evaluate any corresponding physical values.
Suppressing the superscript $''$ again for convenience, one can
write it as
\begin{equation}
{ g^2 \over \sqrt{1+g^2} }=  {1 \over r^2} \label{10}
\end{equation}

Note that the dimensionless physical coordinate $r'' \equiv {r /
r_c}={ r / (\sqrt{m(r)}r_0)}$ is in fact a function of $r$ for a
system with total mass distribution given by $m(r)$. Gauss law
implies that, for a spherically symmetric system, exterior mass
throughout $r>r_1$ will not affect the gravitational field
$g(r<r_1)$. From now on, one will assume $m$ is a constant
independent of the coordinate $r$ for simplicity keeping in mind
that the result is valid only for the exterior region of a
spherical system.

Once $m(r)$ is not a constant, the generalization is still
straightforward. One can simply take $r''$ as co-moving
coordinate. All physics can be re-derived by a proper coordinate
transformation.

Note that Eq. (\ref{10}) can be solved directly to write $g(r)$ as
a function of $r$:
\begin{equation} \label{11}
g= {\sqrt{ 1+ \sqrt{ 1+4r^4} } \over  \sqrt{2} r^2}.
\end{equation}

It is apparent that $g(r)$ goes like $1/r^2$ at short distance
scale where $r \ll 1$. On the other hand, $g(r)$ goes like $1/r$
at large distance scale where $r \gg 1$\cite{kao05}.

\section{Effective Potential of MOND}

In order to take a close look at the changing pattern of $g$, one
can split it into two different parts:\cite{kao05}
\begin{equation}
g=g_l+g_s= { 2\sqrt{2} r^2  \over  \sqrt{1+4r^4} \sqrt{ 1+ \sqrt{
1+4r^4} } } + { \sqrt{ 1+ \sqrt{ 1+4r^4} }  \over  \sqrt{2} r^2
 \sqrt{1+4r^4} }.
\end{equation}
Note that the first term is $g_l$ that goes like $1/r$ when $r \gg
1$, while the second term is $g_s$ that goes like $1/r^2$ when $r
\ll 1$. Therefore it is easy to see that $g_l$ and $g_s$
represents the long distance and short distance field strength of
the $g$ respectively. One is hoping that successful separation of
$g$ may help shedding light to the search of the underlying
theory.

One can integrate $g$ for the effective potential.  After some
algebra, one can show that the effective potential $V_l \equiv -
\int_\infty^r {\bf g}_l \cdot d{\bf r} = \int^r g_l dr$ and $V_s
\equiv \int^r g_s dr$ can be evaluated directly to give
\begin{equation} \label{13}
V_l=  \ln { \sqrt{1+ \sqrt{1+4r^4}} \over \sqrt{2}}-
\sum_{n=1}^{\infty} { \pi_{k=1}^n (4k-3)  \over 2n \cdot n! 2^n
(1+\sqrt{ 1+4r^4} )^n } ,
\end{equation}
and
\begin{equation}\label{14}
V_s =  - { \sqrt{ 1+ \sqrt{ 1+4r^4} } \over\sqrt{2} r }.
\end{equation}
One can verify directly that $V_l'=g_l$ and $V_s'=g_s$ and prove
that above equations are indeed correct up to an irrelevant
integration constant.

In fact, it is difficult to specify this constant of integration
in the conventional approach which take $V(r \to \infty) \to 0$.
This is because the effective potential in fact diverges at
spatial infinity due to the logarithm behavior of the dominating
2D-like potential.

One remarks that the potential $V(r)$ derived here remain valid in
the co-moving coordinate chosen as $r/r_c$ which depends on the
mass content of the spherically symmetric system.

\section{Induced gravity theory and MOND}
One is looking for a theory that goes to Newtonian theory in the
small $r$ region and reproduce the $g \sim 1/r$  effect in the
large $r$ region. The dynamics proposed by MOND gives a specific
force law shown earlier in section II. In particular, the induced
gravity (or equivalently the Brans-Dicke theory) provides a nice
framework for our purpose. Note that induced gravity model,
similar to the Brans-Dicke theory, proposes that the the
gravitational constant is a dynamical variable given by the vacuum
expectation value of a scalar field $\phi$ derivable from the
Lagrangian \cite{w72}
\begin{equation} \label{action}
{\cal L}_\phi=  -{1 \over 8 \omega}\phi  R - {1 \over 8\phi}
(\partial \phi)^2 -W(\phi) .
\end{equation}
Here $R$ is the scalar curvature and $\phi$ is a scalar field
producing a space-time dependent gravitational constant. In
addition, $W(\phi)$ is a spontaneously symmetry broken effective
potential coupled to the system. In addition to the Lagrangian of
scalar field ${\cal L}_\phi$, there is another matter Lagrangian
\begin{equation}
{\cal L}_M= {\cal L}_M(m, \psi)
\end{equation}
with a modified function $\psi(x)$ introduced here to induce the
MOND effect. One will assume that the variation of ${\cal L}^M $
with respect to the metric $g_{\mu \nu}$ will contribute a term
$\psi T^M_{\mu \nu}$ in the field equation accounting for the
matter effect. Here $\psi$ is treated as an auxiliary field
without any dynamics. The main purpose of this auxiliary field is
to introduce a distortion of Newtonian potential in the Newtonian
limit.

The field equations of the induced gravity theory are known to be
\cite{w72}
\begin{eqnarray} \label{eq1}
\phi G_{\mu \nu} &=&-8 \pi \psi T^M_{\mu \nu}  - {\omega \over
\phi}(D_\mu \phi D_\nu \phi - {1 \over 2} g_{\mu \nu} D_\gamma
\phi D^\gamma \phi)\nonumber
\\
 &-&(D_\mu D_\nu \phi -  g_{\mu \nu} D^2 \phi )
+{4 \omega } g_{\mu \nu} W \\ D^2 \phi &=& { 8 \pi \over 3+2
\omega} \psi {T^M_\mu}^\mu + { 8 \omega \over 3+2 \omega} [\phi
\partial_\phi W -2W] \label{eq2}
\end{eqnarray}
with $T^M_{\mu \nu}$ the energy momentum tensor associated with
the matter and $\psi$ another scalar field coupled to the mass
term of the constituent matter.  Note that the variation equation
of $\phi$ will produce a term proportional to the scalar curvature
$R$. This $R$ proportional term can be eliminated by the $R$ term
obtained from taking the trace of the Eq. (\ref{eq1}).  The final
result is exactly the Eq. (\ref{eq2}) shown above.

{\bf Physical interpretation of $\psi$ field:}

Note that mass of a fermion is induced from a Yukawa coupling
term, e.g. $\bar{b}\psi'b$ for a baryon $b$. The mass of the
baryon $b$ is thus induced by the vacuum expectation value of the
coupled scalar field $<\hat{\psi}'> \to m_b$. Assuming that $m_b
(x) =<\hat{\psi}'(x)>$ is spatial dependent, one can in principle
induce a very different mass effect for the system in coherent to
the spatial dependent scalar field $\phi(x)$ introduced earlier.
To be more specific, the inhomogeneous $m_b(x)$ represents the
local mass deformation in our approach.

The overall effect can be integrated in order to produce global
deformation on the Newtonian potential $V(x, m)$. In our approach,
one assumes that the deformation function $\psi(r)$ coupled to the
matter energy-momentum tensor represents collectively the total
effect of the inhomogeneity distortion of MOND potential $V(x)$.
In other words, $\psi$ is assumed to provide the collective
deformation of the entire system accommodating the MOND potential
as a physical resolution. In addition, one assumes that the mass
generating scalar field $\psi'$ is an auxiliary field without a
kinetic term. Moreover, one would like to impose the equation
\begin{equation} \label{22}
\psi V_0=\phi V
\end{equation}
as the auxiliary constraint for $\psi$. Note that what one imposes
here is a relation between $\psi$ and $V$ instead of relating
$\psi$ and $\phi$. Indeed, one can see from the structure of the
field equation (\ref{eq1}) that only the ratio $\psi/\phi$ coupled
to $T^M_{\mu \nu}$ matters. Indeed the field equation reads
$G_{\mu \nu}=[\psi/\phi] T^M_{\mu \nu}+ \cdots $ with the term
$[\psi/\phi] T^M_{\mu \nu}$ serves as the generalized energy
momentum tensor. Therefore, one can parameterize $\psi$
differently, but the final result will remain the same. By all
means, one finds that the constraint (\ref{22}) is the best way to
relate $\psi$ and $V$. This constraint introduced here is similar
to the effect of modified inertial introduced by the scalar-tensor
theory \cite{bek}. By all means, it requires a constraint to be
imposed by hand in order to reproduce the dynamics of MOND in an
exact form. Hopefully, the approach shown here will shed light to
the finding of a more realistic approach.

In fact, the constraint one introduced here is equivalent to
setting $\psi/\phi =V/V_0$ such that the system will become
Newtonian in the limit $V \to V_0$. To be more specific, the
constraint is in fact introduced via the ration of $\phi/\psi$.
Bring $\psi/\phi$ to the right hand side of above equations, one
can write $G_{\mu \nu}=[\psi/\phi] T^M_{\mu \nu}+ \cdots $ as
$\phi' G_{\mu \nu}\equiv [\phi/\psi] G_{\mu \nu}= T^M_{\mu \nu}+
\cdots $. This is equivalent to the induced gravity theory without
a $\psi$ field. One can introduce the constraint via a re-scaling
of the scalar field $\phi'=\phi/\psi$ in the same induced gravity
theory. In this approach, one can as well say that the constraint
is imposed on the choice of scalar field $\phi'=\phi/\psi$ by
choosing a proper gauge of scale transformation \cite{cheng}. And
hopefully, the constraint $\phi'= V_0/V$ (equivalent to the
constraint (\ref{22})) governing the deformation information of
the Newtonian force law will also lead one from Newtonian theory
to the dynamics proposed by MOND. In short, one can either view
the constraint (\ref{22}) as the collective mass function $\psi$
or the deformation of the Newtonian constant $G$ prescribed by
$\phi'$.

Note however that the field equations for these two different
approaches will be slightly different sue to the kinetic term of
the scaled $phi$ field. In a moment we will try to solve this
model in the limit where kinetic term of $\phi$ is omitted. The
field equations of the scalar field will be the same in this
limit. Therefore, it does not matter which viewpoints one decide
to take in the toy model we will study later. The view of the
collective mass deformation is however a quite promising idea,
therefore, one will stick to Eq.s (\ref{eq1},\ref{eq2}) throughout
this paper. In short, the auxiliary field $\psi$ introduced here
is not a dynamical field. In addition, the effect of $\psi$ has
already been absorbed into the collective mass term included in
$\psi T_{\mu \nu}$.

The classical Newtonian field equation can be obtained from the
time-time component of the Einstein equation in the Newtonian
limit with the metric identification
\begin{equation}
g_{00} = -1 -2V.
\end{equation}
Indeed, the geodesic equation
\begin{equation}
{d^2 x^\mu \over ds^2} + \Gamma^\mu_{\alpha \beta} {d x^\alpha
\over ds}{d x^\beta \over ds} =0
\end{equation}
becomes
\begin{equation} \label{20}
{\ddot r} = -{ \partial V \over \partial r}
\end{equation}
in the Newtonian limit. Note that one can write $V =
\varphi(r)V_0$ for convenience with $\nabla^2 V_0=4 \pi \rho$ the
Newtonian potential for a spherically symmetric system. Therefore
$\varphi=V/V_0$ represents the deformation factor signifying the
deviation of the gravitational potential $V$ as compared to the
Newtonian potential $V_0$. One would like to study if there exists
a consistent solution in an induced gravity model that will
accommodate the gravitational potential of the form
$V(r)=(V_l+V_s)$ given by Eq. (\ref{13}-\ref{14}).

Note that the time-time component of the Einstein equation can be
shown to be
\begin{eqnarray}
\label{21} && ({1 \over 2}+V )[\phi''+ {2 \over r} \phi'] -[ V_0
(\varphi'' + {2 \over r} \varphi'
) +2 V_0' \varphi']\phi -\phi \varphi \nabla^2 V_0 \nonumber \\
&&  + 4 \pi \psi \rho = -{4 \omega} ({1 \over 2}+V ) W -\omega ({1
\over 2}+V ) { \phi'^2 \over 2 \phi}
\end{eqnarray}
with $\omega$ a coupling constant.

Since one has $\nabla^2 V_0=4 \pi \rho$, the constraint $\psi
V_0=\phi V$ implying $\psi = \phi \varphi$ can be used to write
Eq. (\ref{21}) as
\begin{eqnarray}
 && ({1 \over 2}+V )[\phi''+ {2 \over r} \phi'] -[ V_0
(\varphi'' + {2 \over r} \varphi'
) +2 V_0' \varphi']\phi \nonumber \\
&& = -{4 \omega} ({1 \over 2}+V ) W -\omega ({1 \over 2}+V ) {
\phi'^2 \over 2 \phi}\label{23}
\end{eqnarray}
In addition, one has
\begin{equation}\label{24}
\phi''+ ({2 \over r}+V') \phi'=  { 8 \pi \over 3+2 \omega} \psi
\rho + { 8 \omega \over 3+2 \omega} [\phi \partial_\phi W -2W]
\end{equation}
from Eq. (\ref{eq1}-\ref{eq2}).

Therefore one has a set of equation with two correlated ODEs and
an unknown scalar potential $W(\phi)$ to be dealt with. In
principle, one can choose some appropriate, possibly exotic,
combination of scalar potential $W$ such that the equations
(23-24) do accommodate consistent solution for $\phi$. One should
first solve the second Eq. (24) to find the solution of $\phi(r)$.
Substituting this solution back to Eq. (23), one should be able to
find $\varphi(r)$.

In practice, what one actually is doing is: (i) given any
prescribed form of Newtonian potential $V(r)$ and insert this
function to equations (\ref{23}-\ref{24}), (ii) one can in
principle solve for a consistent set of solution $(\phi(r),
W(r))$. Finally, one can reconstruct $W(r)$ as a function of
$\phi(r)$. This in principle will give us a solution for
$W(\phi)$. One may however end up with a very complicate
expression for the scalar potential $W$.

To be more specifically, one can write the scalar potential as
\begin{eqnarray}
W &= & \left [ - ({1 \over 2}+V )[\phi''+ {2 \over r} \phi'] + [
V_0 (\varphi'' + {2 \over r} \varphi' ) +2 V_0' \varphi']\phi
\right ]/ \left [ {4 \omega} ({1 \over 2}+V ) \right ]  - {
\phi'^2 \over 8 \phi} \label{23w}
\end{eqnarray}
from Eq. (\ref{23}). In addition, one can put Eq. (\ref{24}) as
\begin{equation}\label{24w}
\phi' \left [ \phi''+ ({2 \over r}+V') \phi' \right ]=  { 8 \pi
\over 3+2 \omega} \psi \phi' \rho + { 8 \omega \over 3+2 \omega}
[\phi W' -2\phi' W]
\end{equation}
by multiplying it with $\phi'$. Eliminating $W'$ and $W$, with the
help of Eq. (\ref{23w}), one can write Eq. (\ref{24w}) as a
differential equation of $\phi$ independent of $W$ with a given
$V(x)$ and $\rho(x)$. This equation can be solved to give a formal
expression of $\phi(r)$. Insert this solution of $\phi(r)$ back to
Eq. (\ref{23w}), one can hence write the scalar potential $W(r)$
as a function of $r$. Since we know the form of $\phi(r)$, one can
invert the function to find $r = r(\phi)$. Therefore, one should
be able to write $W(\phi(r))$ as a functional of $\phi(r)$
straightforwardly. One will show later as an example that the
process does work in the large $r$ limit where the gravitational
force become dramatically deformed to the MOND limit $g \sim 1/r$.

For our purpose, one needs to know if the set of field equations
accommodates arbitrarily specified $\varphi(r)$ as a consistent
solution with a properly chosen $W$. One has two variables $\phi$
and $W(\phi)$ at our disposal. Therefore, the answer turns out to
be yes. Indeed, one can always solve $\phi(\varphi)$ as a function
of $\varphi$ with any given $\varphi(r)$ and $W(\phi)$ from Eq.
(23). The additional Eq. (24) would then require a very special
form of $W$ in order to have the needed function
$\phi(\varphi(r))$ as a consistent solution.

Since consistent solution with any given $\varphi(r)$ can be in
principle made possible with certain properly chosen potential
coupled to the system. Given a carefully chosen potential $W$, one
is naturally lead to the desired deformation factor $\varphi(r)$
that induce MOND as an alternative theory via the identity
$V(r)=\varphi(r)V_0(r)$.

Note that one assumes that the scalar potential $W$ could be of
very complicated origin. It might have to do with some complicated
gravitational field interaction similar to the
temperature-dependent effective potential. One is unable to offer
a solution here. Hopefully the approach shown here may shed a
light to a better understanding of the physical origin of MOND.

For a simple demonstration, one will show explicitly in next
section how to obtain a desired asymptotic solution with a toy
model.

\section{A Toy Model}

One will try to show that the system of equations
(\ref{23}-\ref{24}) does accommodate consistent set of solution
for $\phi$ and $W(\phi)$ in the large $r$ region in this section
with the help of a simple toy model. For simplicity, one will
consider the limit $\omega \to 0$. In order to keep the effective
potential term $W_1$ in our toy model, one also has to assume that
$W_1 = W/\omega $. Equivalently, one is trying to remove the
effect of the kinetic term from the field equations such that the
scalar field $\phi$ becomes an auxiliary field without dynamical
term. To be more specific, one is considering the scalar
Lagrangian given by
\begin{equation} \label{action1}
{\cal L}_\phi=  -{1 \over 8 }\phi  R  -W(\phi) .
\end{equation}
We will try to find models which enable a MOND solution in large
$r$ region where no matter is present. In this case, the Newtonian
potential $V_0 \to 1/r$ according to the Gauss law.

In constructing the model, what one actually did is that: (i) let
$V \to \ln (r/r_1)$ be an approximate solution in this limit, (ii)
insert this solution to equation (\ref{23}-\ref{24}) and solve for
a consistent set of solution $(\phi(r), W(r))$. Finally, one can
reconstruct $W(r)$ as a function of $\phi(r)$.

\subsection{weakly coupled $V$}

In the region $V \ll 1$, $W \ll \phi''$ and $\phi W' \ll \phi''$,
the field equations become:
\begin{eqnarray}
\label{31} && \phi''+ {2 \over r}  \phi' \sim  2 V_0\varphi'' \phi  \\
\label{32} && \phi''+ ({2 \over r}+V') \phi' \sim  0
\end{eqnarray}
with a potential $W$ is either $0$ or close nothing in this
region. Note that one has used the identification $V=\varphi V_0$
and $V_0= -1/r$ in this limit.

It is straightforward to show that the following solutions
\begin{eqnarray}
\label{eqphi2}  \phi & \to &  {\phi_0 \over r^2} \\
V & \to & \ln {r \over r_1}
\end{eqnarray}
solve Eq. (\ref{31}-\ref{32}) in a consistent way. Here the
parameter $r_1$ represents a local re-scaling of the zero point of
the gravitational potential. The constant $ -\ln r_1$ will not
affect the force law Eq. (\ref{20}).

In addition, the constraint $V \ll 1$ implies that this solution
is good for $r$ close to $r_1$. For $r_1 \sim r_0$, the solution
does reproduce the $1/r$ force law beyond the luminous region of
spiral galaxies. It was shown in Ref. \cite{kao05} that $V \to \ln
r$ quickly when $r > r_c$. Hence region close to $r_c$ is a
perfect domain for our assumption.

Note that the assumption $W \ll \phi''$ and $\phi W' \ll \phi''$
implies that the scalar potential $W(\phi)=W(\phi_0/r^2) \ll
\phi'' \sim 1/r^4$ and $W'(\phi)=W'(\phi_0/r^2) \ll 1/r^4$ in this
region. For example, $W = \lambda \phi^4$ is a perfect candidate
for the scalar potential near the region $r \sim r_1$. If $r_1$ is
bigger than the galactic size, this solution provides a reasonable
and good prediction for many RC observations.

One knows that the scalar potential $W$ may be affected by the
detailed dynamics of the system. For example, it will depend on
the temperature of the system \cite{temp}.  Therefore, it is also
natural to expect a potential correlated to the scale length $r$
in a large-scale astrophysical system. Once the effective scalar
potential becomes negligible in a domain near and beyond our
galactic size, the toy model discussed here may provide a good
resolution for the new dynamics in that region.

\subsection{large $r$ region}

One will show that if the scalar potential is given by
\begin{equation} \label{W1}
W \sim  -{3 \phi_0^2 \over 8 \phi} \exp [- 2 \phi/\phi_0]
\end{equation}
in the large $r$ region, one will be able to reproduce a
gravitational potential $V \to \ln r$ asymptotically in the large
$r$ region.

Note that we are assuming that the weak field approximation
$g_{00} \to -1-2V$ and the resulting field equations (23-24) still
hold in the large $r$ limit even the assumption $V \ll 1$ is not
valid in the large $r$ limit.

There are two main reasons for this approach. Firstly, MOND
approach also have similar problem with the asymptotic divergent
potential $V$ that goes like $\ln r$ in the large $r$ region. One
has to deal with the potential difficult region in any case. If
the induced gravity model considered here has anything to with the
real physics, there must, hopefully, exist certain sort of
normalization process, similar to the re-normalization theory,
holding Eq. (\ref{23}-\ref{24}) valid in the asymptotic region.
Secondly, the purpose dealing with this toy model is simply to
demonstrate how it is possible to derive the desired large scale
(or weak field) gravitational potential $V$ consistently, at least
in the large $r$ region.

The field equations become
\begin{eqnarray}
\label{27} && V \phi''+ {2 \over r}V  \phi' \sim  [
 V_0
(\varphi'' + {2 \over r} \varphi') +2 V_0' \varphi' ] \phi -4VW \\
&& \phi''+ ({2 \over r}+V') \phi' \sim { 8 \over 3} [\phi \partial_\phi W -2W] \nonumber \\
  &&\sim 2 \phi_0 \exp [- 2
\phi/\phi_0] - 8 W. \label{28}
\end{eqnarray}
It is easy to show that Eq. (\ref{28}) does have an asymptotic
approximate-solution $\phi \to \phi_0 \ln r$ by ignoring the
negligible terms proportional to $W$. One can readily show that
the $W$ term is indeed negligible when $ \phi
 \to \phi_0 \ln r$ in Eq. (\ref{28}). Substituting this asymptotic solution $\phi$ into Eq.
(\ref{27}), one can derive the following equation
\begin{equation}
\ln r \; \varphi''  \sim {1 \over r^2} \varphi
\end{equation}
Note that one has also ignored the negligible $W$-dependent term
in Eq. (\ref{27}). One can hence easily show that the solution to
above equation is
\begin{equation}
\varphi \to - r \ln r \label{varphi_lr}
\end{equation}
which implies the asymptotic solution $V \to \ln r$ in the large
$r$ region.

Therefore, one shows that it is indeed possible to derive the
asymptotic solution with a toy model given by the action
(\ref{action1}). In principle, it is possible to reproduce any
physical deformation of the gravitational potential one desires
with the help of some properly chosen scalar potential.

Note that one in fact expects that $V \to \ln r$ and hence
$\varphi \to - r \ln r$ as give by Eq. (\ref{varphi_lr}).
Therefore, one inserts this solution back to the field equation
and try to find potential $W$ which works along with the field
equations. As a result, one finds that $W \sim  -3 \phi_0^2 \exp
[- 2 \phi/\phi_0]/8 \phi$ is the effective potential one is
looking for.

\section{conclusion}
The force law of MOND is reviewed and rewritten in term of
dimensionless field variables for a spherically symmetric system
in this paper.

One of the effect of the scalar field is to deform the effective
distance between two distant matters. The scalar field $\psi$
coupled to the matter energy momentum tensor is proposed to
represent the spatial inhomogeneous effect of the distance
deformation in a collective way. One has also shown explicitly in
this paper a possible resolution derived from a simple model
incorporated with a spherically symmetric mass-distribution
generated by similar effect.

Indeed, one shows that the dynamics of MOND could be due to the
large-scale spatial inhomogeneity of the coupled scalar field. A
specific example with a toy model is presented in this paper to
demonstrate its effect in two different regions: (i) one presents
an approximate solution in the limit $V \sim \ln (r/r_1) \ll 1$
which holds in the region near $r=r_1$. (ii) approximate solution
is obtained in the large $r$ limit. Note that the effective
potential $V(r)$ goes like $\ln r$ in the absence of any dark
matter in this toy model. Indeed,

 Eq.s (\ref{23}-\ref{24}) are in principle solvable with some
properly chosen potential even one is unable to obtain an
analytical solution at the moment. The method shown in this paper
also applies to different force law, with a different
$\varphi(r)$, that may properly describe the RC with or without
any sort of dark matter. The constraint (\ref{22}) is introduced
as an auxiliary constraint. One does not know the physical origin
of this constraint other than the consistency in dimension. It
deserves more attention studying the physical origin of this
constraint.

 \vskip 0.5cm

 {\bf \large Acknowledgments} This work is supported in
part by the National Science Council of Taiwan.

\end{document}